\documentstyle[aps,prd]{revtex}
\tighten
\renewcommand{\epsilon}{\varepsilon}
\begin{document}

\title{Cosmological particle production, causal thermodynamics, and inflationary expansion}

\author{Winfried Zimdahl\footnote{Electronic address: winfried.zimdahl@uni-konstanz.de}}

\address{Fakult\"at f\"ur Physik, Universit\"at Konstanz, PF 5560 M678
D-78457 Konstanz, Germany}

\date{\today}

\maketitle

\pacs{98.80.Hw, 04.40.Nr, 95.30.Tg, 05.70.Ln}

\begin{abstract}
Combining the equivalence between cosmological particle creation and an  effective viscous fluid pressure with the fact that the latter represents a dynamical degree of freedom within the second-order Israel-Stewart theory for imperfect fluids, we reconsider the possibility of accelerated expansion in fluid cosmology. 
We find an inherent self-limitation for the magnitude of an effective bulk pressure which is due to adiabatic (isentropic) particle production. 
For a production rate which depends quadratically on the Hubble rate we confirm the existence of solutions which describe a smooth transition from inflationary to noninflationary behavior and discuss their interpretation within the model of a decaying vacuum energy density. 
An alternative formulation of the effective imperfect fluid dynamics in terms of a minimally coupled scalar field is given. The corresponding potential  
is discussed and an entropy equivalent for the scalar field is found. 
\end{abstract}

\section{Introduction}
It is the essential feature of the second-order thermodynamical theories of M\"uller \cite{Mue}, Israel and Stewart \cite{I,ISt,HL}, and Pav\'on et al. \cite{PJC,PBJ} that the bulk and shear viscous pressures and the heat flux, which describe the deviations of a relativistic fluid from local equilibrium, become dynamical degrees of freedom on their own and obey ``causal'' evolution equations guaranteeing subluminal propagation speeds of thermal and viscous perturbations. 
By now, it is generally agreed that these second-order theories should replace the traditional first-order theories by Eckart and Landau and Lifshitz, which suffer from serious drawbacks concerning causality and stability. 

In homogeneous and isotropic models of the universe a bulk viscous pressure is the only possible dissipative phenomenon. 
The role of bulk pressures in  cosmic media is twofold. 
Firstly, they generally occur if different components of the cosmic substratum are coupled. 
Due to their different internal equations of state the cooling of the subfluids with the expansion of the universe is different, resulting in a tendency of the system to move away from equilibrium.  
These different cooling rates of the components give rise to a bulk viscous pressure of the cosmic medium as a whole \cite{Wein,Strau,Schw,UI,ZMN96}. 
Secondly, a nonvanishing bulk pressure of the cosmic fluid may be the consequence of particle number nonconserving interactions. In particular, 
one may think here of quantum particle production out of the gravitational field \cite{Zel,Mur,Hu}. 
The manifestation of cosmological particle creation as an effective bulk pressure has initiated a series of papers discussing a fluid phenomenological approach to cosmological particle production 
\cite{Turok,Barr,Prig,Calv,LiGer,ZP1,ZP2,ZP3,GaLeDe,Lima,ZPRD,TZP,ZTP,ZMN97,GunzMaNe,ZGeq,ZiBa1,ZiBa2,JeGer}. 
The present paper will mainly explore this second aspect as well. 

Special attention has been devoted to ``isentropic'' (or ``adiabatic'') particle production \cite{Prig,Calv}, i.e., production of perfect fluid particles. 
Here ``isentropic'' means constant entropy per particle. 
There is, however, entropy production due to the enlargement of the phase space of the system since the number of perfect fluid particles increases. 
The condition of isentropic particle production establishes a simple relationship between the particle production rate and the viscous pressure which in most application up to now has been investigated in connection with an Eckart-type expression for the entropy flow vector (see, however, 
\cite{GaLeDe,ZiGaLeDe}). 

Because of its generally negative sign many authors have addressed the question whether a cosmological bulk pressure may become sufficiently large to induce a phase of accelerated expansion (  \cite{Diosi,Pach,LiPoWa,HiSa,ZaJ,RoyCQG,RM,GaLeDe,ZPRD}). 
The basic shortcoming of most previous attempts to establish scenarios of bulk-viscosity-driven inflation for cosmic media with conserved particle number is an apparently unphysical behavior of the thermodynamic quantities. 
The point is that the non-equilibrium viscous pressure has to be of the order of or even to exceed the equilibrium pressure. 
This assumption is in itself problematic and difficult to justify. 
Moreover, as a result of the back reaction of a viscous pressure of this order on the temperature, the universe heats up while at the same time the cosmic matter is diluted \cite{ZPRD}. 
It  is difficult to imagine interaction processes with these consequences in particular in the extreme case of an exponential expansion 
\cite{RoyCQG,ZPRD,RM}. 
In this respect the causal second-order theory is not really superior to the traditional Eckart theory. 

The situation is different, however, if the fluid particle number is allowed to change, i.e., if cosmological particle production processes are taken into account. 
The particle production rate may be regarded as an additional parameter for the fluid dynamics. 
Under such conditions the corresponding effective bulk pressure is not necessarily small. 
For example, the production rate may be fixed by symmetry requirements, corresponding to specific self-interactions of the cosmic medium \cite{ZiBa1,ZiBa2}. 

In the present paper we follow a different strategy. We combine cosmological particle production with an entropy-flow expression of the Israel-Stewart type. 
However, our considerations continue to rely on the concept of isentropic particle production and therefore retain the simple relationship between creation rate and viscous pressure, mentioned above. 
This procedure allows us to replace the bulk pressure in the second-order evolution equation by the corresponding particle production rate and to make the latter quantity an independent degree of freedom. 
The particle production rate is dynamized and no longer a free parameter.  
This provides the framework in which we reconsider the question whether phases of inflationary expansion may be consistent with a fluid cosmological description. 
It turns out that this setting avoids the drawbacks of ``bulk viscous inflation'' without particle production. 
We establish a model where the universe starts in a de Sitter phase with finite, stationary values of density and temperature, and subsequently evolves smoothly to a standard FLRW stage. 
We point out similarities and differences to the scenario by Gunzig et al. \cite{GunzMaNe}, 
which is based on the assumption of a decaying cosmological ``constant''. 
In particular, within the causal, second-order theory the corresponding 
``deflationary'' \cite{Barrow} solution is subject to additional thermodynamical constraints which represent a self-limitation concerning the admissible magnitude of the effective viscous pressure.  
 
Moreover, we clarify how the effective imperfect fluid picture is related to the dynamics of a scalar field. 
While the scalar-field-based inflationary dynamics relies on  a ``slow roll'' period with ``adiabatic supercooling'', followed by a ``reheating'' phase during which all the entropy is produced, the fluid picture implies 
entropy production already during the de Sitter phase. 
The interrelation between the fluid and the scalar field pictures will allow us to express the growing entropy (due to particle production) in a comoving volume also in terms of the scalar field variable.

The paper is organized as follows. 
In Sec. II we summarize (and somewhat generalize) the second-order theory of a homogeneous and isotropic, spatially flat bulk viscous universe and point out the problems inherent in ``bulk viscous inflation'' without particle production. 
Section III is devoted to the dynamics of a universe with ``isentropic'' particle production on the basis of a second-order expression for the entropy flow vector.  The creation rate is regarded as a dynamical degree of freedom. 
We consider linear and quadratic dependences of the particle production rate on the Hubble parameter and discuss the self-restriction of this rate as a consequence of the positivity requirement for the corresponding relaxation time. 
For the case of a quadratic dependence an inflationary model of the universe with a smooth transition to a Friedmann-Lema\^{\i}tre-Robertson-Walker 
(FLRW) stage is established and its relation to a scenario with decaying cosmological term is discussed. 
The relation between fluid cosmology with isentropic particle production and an alternative scalar field description is clarified in Sec. IV. In Sec. V we summarize our results. 
Units have been chosen so that $c = k_{B} =  \hbar =1$. 

\section{The viscous universe}
We assume the cosmic matter to be characterized by the energy momentum tensor of a bulk viscous fluid, 
\begin{equation}
T^{ik} = \rho u^{i}u^{k} 
+ \left(p + \pi\right) h^{ik} \ ,
\label{1}
\end{equation}
and the particle flow vector
\begin{equation}
N ^{i} = n u ^{i}\ .
\label{2}
\end{equation}
Here, $\rho $ is the energy density, $p$ is the equilibrium pressure, 
$n$ is the particle number density, $u ^{i}$ is the fluid four-velocity in the Eckart frame, and $h ^{ik} = g ^{ik} + u ^{i}u ^{k}$ is the spatial projection tensor. 
The quantity $\pi $ denotes that part of the scalar pressure which is connected with entropy production. 
The conservation laws  
$N^{a}_{;a}  =  0$ and $T^{ab}_{\ ;b}  =  0$ then imply   
\begin{equation}
\dot{n} + \Theta n =  0  
\label{3}
\end{equation}
and
\begin{equation}
\dot{\rho} = - \Theta\left(\rho + p + \pi\right)\ ,
\label{4}
\end{equation}
respectively. 
Taking into account second-order deviations from equilibrium, the
entropy flow vector $S^{a}$ takes the form \cite{IS}
\begin{equation}
S^{a} = sN^{a} 
- \frac{\tau\pi^2}{2\zeta T} u^{a} \ , 
\label{5}
\end{equation}
where $s$ is the entropy per particle, $T$ is the fluid temperature, and 
$\tau$ is a  relaxation time.  
The coefficient of  bulk
viscosity is denoted by $\zeta$. 
For the entropy production density one obtains  
\begin{equation}
T S ^{m}_{;m} = - \pi \left[\Theta + \frac{\tau }{\zeta }\dot{\pi } + 
\frac{T}{2}\left(\frac{\tau }{\zeta T} u ^{m}\right)_{;m}\pi  \right]\ .
\label{6}
\end{equation}
The simplest way to guarantee  $S^{a}_{;a} \geq 0$ is a generalized linear relation 
\begin{equation}
\pi = - \zeta \left[\Theta + \frac{\tau }{\zeta }\dot{\pi } + 
\frac{T}{2}\left(\frac{\tau }{\zeta T} u ^{m}\right)_{;m}\pi  \right]\ ,
\label{7}
\end{equation}
which implies the evolution
(see, e.g., \cite{ZPRD} and references therein)
equation
\begin{equation}
\pi + \tau\dot{\pi}  =  - \zeta \Theta - 
\frac{1}{2}\pi \tau\left[\Theta + \frac{\dot{\tau}}{\tau} 
- \frac{\dot{\zeta}}{\zeta} - \frac{\dot{T}}{T} 
\right] 
\label{8}
\end{equation}
for $\pi$, as well as the expression 
\begin{equation}
S^{a}_{;a}  =  \frac{\pi^{2}}{\zeta T}
\label{9}
\end{equation}
for the entropy production density. 
It is the basic feature of the causal non-equilibrium theory that the viscous pressure $\pi $ becomes a dynamical degree of freedom on its own and is subject to an evolution equation (\ref{8}) .  
For $\tau \rightarrow 0$ Eq.(\ref{8}) reduces to the corresponding algebraic relation $\pi _{E} = -\zeta \Theta $ of
the Eckart theory (subscript E). 

We will assume the equilibrium variables $p$ and $\rho $ to obey equations of state 
\begin{equation}
p = p \left(n,T \right)\ ,\ \ \ \ \ \ \ \ \ 
\rho = \rho \left(n,T \right)\ .
\label{10}
\end{equation}
The temperature behaviour of a bulk viscous fluid is governed by 
\cite{Calv,LiGer,ZPRD,RM}
\begin{equation}
\frac{\dot{T}}{T}  = - \Theta
\left[\frac{\partial p}{\partial \rho} 
+  \frac{\pi}{T \partial \rho/\partial T}\right]\ ,   
\label{11}
\end{equation}
where 
\[
\frac{\partial{p}}{\partial{\rho }} \equiv  
\frac{\left(\partial p/ \partial T \right)_{n}}
{\left(\partial \rho / \partial T \right)_{n}}\ ,
\ \ \ \ \ \ \ \ \ 
\frac{\partial{\rho }}{\partial{T}} \equiv  
\left(\frac{\partial{\rho }}{\partial{T}} \right)_{n} \ ,
\ \ \ \ 
{\rm etc.}
\]
For $\pi = 0$ the temperature law (\ref{11})  reproduces
the well known $T \propto a^{-1}$ behaviour in a radiation 
dominated FLRW universe, 
while we recover  
$T \propto a^{-2}$ in the matter dominated case. 
For a viscous fluid  the behaviour of the
temperature depends on $\pi$. 
Since $\pi$ is expected to be negative, the second term in the
bracket on the right-hand side of Eq. (\ref{11}) will counteract the first one.  The existence of a
nonvanishing 
bulk viscous pressure implies that the temperature decreases 
less than  for a perfect fluid. 
$\pi$ tends to `reheat' the cosmic matter. 

Use of the equations of state (\ref{10}) and the temperature law 
(\ref{11}) allows us to write the evolution equation (\ref{8}) for the viscous pressure as 
\begin{equation}
\pi + \tau\dot{\pi}  =  
- \rho\Theta\tau\left[\gamma c _{b}^{2} +  
\frac{\pi}{2\rho}
\left(2 + \frac{\partial{p}}{\partial{\rho }}
+ c _{s}^{2} \right)   
+ \frac{\pi^{2}}{2 \gamma \rho^{2}} 
\left(\frac{\rho + p}{T \partial \rho / \partial T} 
+ 1 + \frac{\partial{p}}{\partial{\rho }}
\right)\right]
+ \frac{\tau \pi }{2}
\frac{\left(c _{b}^{2}\right)^{\displaystyle \cdot}}
{c _{b}^{2} }\ ,
\label{12}
\end{equation}
where 
\begin{equation}
c _{s}^{2}  = \left(\frac{\partial{p}}{\partial{\rho }} \right)_{ad} 
= \frac{n}{\rho + p}\frac{\partial{p}}{\partial{n}} 
+ \frac{T}{\rho + p} 
\frac{\left(\partial p / \partial T \right)^{2}}
{\partial \rho / \partial T}\ 
\label{13}
\end{equation}
is the square of the adiabatic  sound velocity $c _{s}$. 
The quantity $c _{b}$ is the  
propagation velocity of viscous pulses, determined by \cite{HL,RM}
\begin{equation}
c _{b}^{2} =  
\frac{\zeta }{\left(\rho + p \right)\tau }\ .
\label{14}
\end{equation}
The sound in a viscous medium propagates with a subluminal velocity $v$ 
where \cite{RM}
\begin{equation}
v ^{2} = c _{s}^{2} + c _{b}^{2}  \leq 1\ .    
\label{15}
\end{equation}
With the help of the field equations ($\kappa$ is Einstein's gravitational constant and $H$ is the Hubble parameter $H \equiv  \Theta /3$)  
\begin{equation}
\kappa \rho = 3 H ^{2}\ ,
\mbox{\ \ \ \ }
\dot{H} = - \frac{\kappa}{2}\left(\rho + p + \pi  \right)
\label{16}
\end{equation} 
for an homogeneous and isotropic, spatially flat universe, we may write  
\begin{equation}
\kappa \pi = - 3 \gamma H ^{2} - 2 \dot{H}\ ,
\label{17} 
\end{equation}
where $\gamma \equiv  1+p/ \rho $,  
and
\begin{equation}
\kappa \dot{\pi } = - 2 \ddot{H} 
- 6 H \dot{H}
\left(1 + \frac{\partial{p}}{\partial{\rho }} \right) 
+ 9 H ^{3}\gamma \left(c _{s}^{2} 
- \frac{\partial{p}}{\partial{\rho }} \right)\ .
\label{18}
\end{equation}
Combining the last two relations with Eq. (\ref{12}), the viscous pressure 
and its derivative may be eliminated to yield the basic 
causal evolution equation for the Hubble parameter:  
\begin{eqnarray}
&&\tau H \left\{\frac{\ddot{H}}{H} 
- \frac{\dot{H}^{2}}{H ^{2}}\gamma ^{-1}
\left(\frac{\rho + p }{T \partial \rho / \partial T} + 1 
+ \frac{\partial{p}}{\partial{\rho }}\right)  
- 3\dot{H}\left[\frac{\rho + p}{T \partial \rho / \partial T} 
- 1 - \frac{\partial{p}}{\partial{\rho }} 
+ \frac{1}{2}\left(\frac{\partial{p}}{\partial{\rho }} 
- c _{s}^{2} \right)
\right]  
\right.
\nonumber\\
&&- \left. \frac{9}{2}H ^{2}\gamma 
\left[c _{b}^{2}   
+ \frac{1}{2}
\left(\frac{\rho + p}{T \partial \rho / \partial T} 
- 1 - \frac{\partial{p}}{\partial{\rho }}   
\right)
- \frac{1}{2}\left(\frac{\partial{p}}{\partial{\rho }} 
- c _{s}^{2} \right)
\right] 
- \frac{1}{2}\frac{\left(c _{b}^{2} \right)^{\displaystyle \cdot}}
{H c _{b}^{2}}\left(\dot{H} + \frac{3}{2}\gamma H ^{2}\right)
\right\} + \dot{H}
+ \frac{3}{2}\gamma H^{2} = 0 \ . 
\label{19}
\end{eqnarray}
This equation is valid for arbitrary, even arbitrary time varying equations of state. 
To get an idea of the physical implications of Eq. (\ref{19}) we introduce the notations
\begin{equation}
r \equiv  \frac{\rho }{T \partial \rho / \partial T} \ , \ 
\gamma A \equiv  \gamma r + 1 + \frac{\partial{p}}{\partial{\rho }}\ , \ 
B \equiv  \gamma r - 1 - \frac{\partial{p}}{\partial{\rho }}\ , \ 
C \equiv  \frac{\partial{p}}{\partial{\rho }} - c _{s}^{2} \ ,  
\label{}
\end{equation}
with $1 \leq \gamma \leq 4/3$ and assume the values of $r$, $A$, $B$, $C$, and $c _{b}$   to be constant. 
In the ultrarelativistic limit with 
$\gamma = 4/3$, $\partial p/ \partial \rho = c _{s}^{2}=1/3$  we have 
$r=1$, $A=2$, and $B=C=0$. 
For non-relativistic matter, characterized by  
$\gamma = 1$, $\partial p/ \partial \rho = 2/3$, $c _{s}^{2}\ll 1$ the corresponding quantities are 
$r\gg 1$, $A \gg 1$, $B \gg 1$, and $C=2/3$.  
The general equation (\ref{19}) then simplifies to
\begin{equation}
\tau H \left\{\frac{\ddot{H}}{H} 
- A\frac{\dot{H}^{2}}{H ^{2}} 
- 3 \dot{H}\left[B + \frac{C}{2}\right] 
- \frac{9}{2}\gamma H ^{2}
\left[c _{b}^{2} + \frac{1}{2}\left(B-C \right) \right]
\right\} + \dot{H}
+ \frac{3}{2}\gamma H^{2} = 0 
\ .
\label{20}
\end{equation}
This equation has stationary solutions $\dot{H} = 0$ for which 
\begin{equation}
\tau H = \frac{1}{3 \left[c _{b}^{2} + \frac{1}{2}\left(B-C \right) \right]}
\label{21}
\end{equation}
is valid. 
Causality restricts $c _{b}$ to values 
$c _{b}^{2}\leq 1 - c _{s}^{2} $ [cf. Eq. (\ref{15})]. 
For equations of state close to that for non-relativistic matter one has 
$c _{b}^{2}\leq 1 $ as well as $B \gg 1$ and $C \approx 2/3$  and the relaxation time $\tau $ is much shorter than the cosmological time scale 
$H ^{-1}$.   
However, for equations of state close to that for radiation with 
$r \approx 1$, $c _{b}^{2} \leq 2/3 $, $B \approx 0$, and 
$C \approx 0$, the relaxation time may well be of the order of the Hubble time, i.e., the nonequilibrium may be ``frozen in''. 
Such kind of ``freezing in'' may be regarded as a necessary condition for successful inflation \cite{RoyCQG} since otherwise the system would relax to equilibrium in less than one Hubble time. 
Since, on the other hand, one expects $\tau $ to be typically of the order of the mean free collision time and this collision time itself has to be smaller than the Hubble time for a fluid description to make sense, we face here serious applicability limits of a viscous fluid approach to inflationary cosmology. 
In fact, the situation is still more unfavorate. The stationary  
case implies $\pi = - \left(\rho + p \right)$ and 
$T \propto a ^{3 \left[\gamma r - \partial p/ \partial \rho  \right]}$ 
with $a \propto \exp{\left[Ht \right]}$ , i.e., the temperature is exponentially increasing for any equation of state. 
The number density in contrast is exponentially decreasing as 
$n \propto a ^{-3}$. This decrease compensates the increase in $T$, resulting in a constant energy density $\rho$. 
A behavior such as this, although formally not excluded, seems highly unrealistic. 

Let us now look for solutions $a \propto t ^{q} $ with constant values of $q$,  equivalent to 
$H = q t ^{-1}$, $\dot{H} = -q t ^{-2}$, and $\ddot{H} = 2 q t ^{-3}$. 
The parameter $\tau H$ becomes 
\begin{equation}
\tau H = \frac{1 - \frac{2}{3 \gamma q}}{3c _{b}^{2}}Q
\ ,
\label{22}
\end{equation}
with the abbreviation 
\[
Q \equiv  \frac{c _{b}^{2}  }
{c _{b}^{2} + \frac{1}{2}\left(B-C \right) 
+ \frac{2}{9 \gamma q ^{2}}\left[A-2-3q \left(B+C \right) \right]}\ .
\]
In the ultrarelativistic limit $Q$ approaches $Q=1$, while 
$Q \ll 1$ in the opposite limit of non-relativistic matter.    
On the other hand, we find 
\begin{equation}
- \frac{\pi }{\rho + p} = 1 - \frac{2}{3 \gamma q}
\quad\Rightarrow\quad 
\tau H = - \frac{1}{3c _{b}^{2}}\frac{\pi }{\rho + p}Q
\ .
\label{23}
\end{equation}
For $\pi \neq 0$ we have $\tau \rightarrow 0$ in the limit 
$c _{b}^{2} \rightarrow \infty$ which is the Eckart case. 
The last equation of Eq. (\ref{23})  represents an algebraic relation 
between the relaxation time $\tau $ and the viscous pressure $\pi $. 
Solving the first relation of Eq. (\ref{23}) for $q$ yields
\begin{equation}
q = \frac{2}{3 \gamma \left[1+ \frac{\pi }{\rho + p} \right]}
\ .
\label{24}
\end{equation}
Eqs. (\ref{23}) and (\ref{24}) imply a constant ratio 
$\pi /\rho$   under the condition of a power-law behaviour of the scale factor. 
The temperature law (\ref{11}) becomes
\begin{equation}
\frac{\dot{T}}{T} = -3 H \left[\frac{\partial{p}}{\partial{\rho }} 
+ \frac{\pi }{\rho + p} \gamma r\right]
\quad\Rightarrow\quad 
T \propto a ^{3 \left[\gamma r \left(1-\frac{2}{3 \gamma q} \right) 
- \partial p/ \partial \rho  \right]}
\ .
\label{25}
\end{equation}
The case $q = 2/ \left(3 \gamma  \right)$ corresponds to $\pi = 0$ (perfect fluid limit). 
Since $Q>0$ only negative values of $\pi $ guarantee a positive relaxation time $\tau $ which is consistent with a positive bulk viscosity coefficient according to Eq. (\ref{14})  and positive entropy production according to Eq. (\ref{9}). 
The case $q = 1$ corresponding to $\ddot{a} = 0$, is characterized by 
$\pi = - \left(1-2/ \left(3 \gamma  \right) \right)
\left(\rho + p \right)$. 
The energy density decays as $\rho \propto a ^{-2}$, i.e., it mimicks the behaviour of so-called ``K-matter'' \cite{Kolb}. 
The temperature increases, however, as 
$T \propto a ^{3 \left[\gamma r \left(1-2/\left(3 \gamma  \right) \right) 
- \partial p/ \partial \rho \right]}$ in this case.  
The number density $n$ decays as $n \propto a ^{-3}$ for arbitrary $q$. 
For any $q > 1$ we have accelerated expansion, i.e., power-law inflation. 
For very large $q$ the viscous pressure $\pi $ approaches 
$- \left(\rho + p \right)$, the case of exponential inflation. 
For any value $q > 2/3$ the non-equilibrium pressure exceeds the equilibrium 
one, i.e. $|\pi|>p $,  and we are beyond the strictly justified range of applicability of the theory.  

The considerations so far summarize and generalize previous presentations of bulk viscous cosmology \cite{ZPRD}. 
They confirm that within the framework presented above, relying on a constant number of particles, the concept of bulk-viscosity-driven inflation is hardly convincing. 
However, particle number conservation is an exceptional situation in early cosmology. 
In the following section we shall extend the theoretical basis of our investigations insofar as we will admit the possibility of a nonvanishing source term on the right-hand side of Eq. (\ref{3}), i.e., particle number conservation will no longer be required.

\section{The universe with particle production}
\subsection{General theory}
In case the fluid particle number is not preserved, the conservation law  (\ref{3}) has to be replaced by the balance \cite{Prig,Calv}
\begin{equation}
N^{a}_{;a}  =  
\dot{n} + \Theta n =  n \Gamma  \ , 
\label{26}
\end{equation}
where $\Gamma = \dot{N}/N$ is the change rate of the number 
$N \equiv  n a ^{3}$  of particles in a comoving volume $a ^{3}$. 
For $\Gamma > 0$ we have particle creation, for $\Gamma < 0$ particles are annihilated. 
A nonvanishing particle production rate is known to generate an effective bulk pressure of the cosmic fluid. 
This is most simply demonstrated for ``isentropic'' particle production. 
From the Gibbs equation 
\begin{equation}
T \mbox{d}s = \mbox{d} \frac{\rho }{n} - p \mbox{d} \frac{1}{n}
\label{27}
\end{equation}
together with the balances (\ref{4}) and (\ref{26})  we obtain 
\begin{equation}
nT \dot{s} = - \Theta \pi - \left(\rho + p \right)\Gamma \ .
\label{28}
\end{equation}
``Isentropic'' particle production is characterized by $\dot{s} = 0$. 
Under this condition 
the equilibrium entropy per particle does not change as it does in dissipative processes of the type discussed in Sec. II. 
Instead, the viscous pressure is entirely determined by the particle production rate \cite{Prig,Calv}: 
\begin{equation}
\dot{s} = 0 \quad\Rightarrow\quad  
\pi = - \left(\rho + p \right)\frac{\Gamma }{\Theta }\ .
\label{29}
\end{equation}
The cosmic substratum is not a conventional dissipative fluid but a perfect fluid with varying particle number. 
There is, however, entropy production due to the enlargement of the phase space of the system. 
First we consider the analogue of the Eckart theory (subscript E). 
From the entropy flow vector 
$S ^{a}_{E}=nsu ^{a}$ one obtains a nonvanishing entropy production although 
$\dot{s}=0$:
\begin{equation}
S ^{a}_{E;a} = ns \Gamma \ .
\label{30}
\end{equation}
While in principle the rate $\Gamma $ has to be calculated from microphysics 
(it may also be prescribed by symmetry requirements connected with specific inelastic self-interactions of the cosmic matter \cite{ZiBa1,ZiBa2}), a phenomenological determination along the lines of linear, irreversible thermodynamics is possible here as well. 
To this purpose we rewrite Eq. (\ref{30}) with condition (\ref{29}) as 
\begin{equation}
S ^{a}_{E;a} = - \frac{nsT}{\rho + p}\frac{\Theta \pi }{T}. 
\label{31}
\end{equation}
(Notice that $nsT= \rho + p$ for a vanishing chemical potential.)  
The simplest way to guarantee  $S ^{a}_{E;a} \geq 0$ is the linear ansatz 
\begin{equation}
\pi _{E} = - \zeta \frac{nsT}{\rho + p}\Theta 
\equiv  - \zeta _{eff}\Theta 
\quad\Rightarrow\quad 
\Gamma _{E} = \frac{\zeta _{eff}}{\rho + p}\Theta ^{2}\ ,
\label{32}
\end{equation}
with a coefficient $\zeta >0$. Of course, this implies $\Gamma _{E} >0$. 
Only a positive $\Gamma _{E}$ is compatible with the second law of thermodynamics. 
The existence of relations of the type (\ref{32}) shows that there is a close analogy to the conventional first-order theory of dissipative processes 
(in which $\Gamma =0$ and $\dot{s}\neq 0$).   
It appears natural to extend this analogy also to the second-order theory 
where the viscous pressure becomes a dynamical degree of freedom.  In the present setting this allows us to dynamize the particle production rate 
$\Gamma $ which was introduced as an additional parametern in the balance (\ref{26}). 
Our approach starts again with the expression (\ref{5}) for the entropy flow vector. However, due to the possibility of a varying particle number,   
Eq. (\ref{6}) is now replaced by  
\begin{equation}
S ^{m}_{;m} = n \dot{s} + s N ^{m}_{;m} 
- \frac{1}{2}\left(\frac{\tau }{\zeta T}u ^{m} \right)_{;m}\pi ^{2} 
- \frac{\tau }{\zeta T}\pi \dot{\pi }\ .
\label{33}
\end{equation}
While the expression (\ref{5}) was derived under the condition of small deviations from the standard perfect fluid behavior, we will not explicitly apply this restriction in the following.   
Instead, we will show that in case $\pi $ is entirely a consequence  of cosmological particle creation, the dynamics based on Eq. (\ref{33}) may provide its own limits. Such kind of self-limitation does not exist for a conventional viscous pressure $\pi $.   
Nevertheless, a really satisfactory theory should be based on a higher-order generalization of the expression (\ref{5}). 
(For a recent attempt in this direction see \cite{MaaMe}). 
The present approach may be regarded as a first step which exploits the fact that the entropy flow (\ref{5}) is the simplest option which allows us to  dynamize the particle creation rate on the phenomenological level at all.   

Focusing on the case  (\ref{29}) of ``isentropic'' particle production, i.e., on processes with constant entropy per particle, means that  
the variable $\pi $ does not characterize a conventional nonequilibrium but a state with equilibrium properties as well, which, however, deviates from the specific equilibrium with $\Gamma =0$. 
The equilibrium condition $\dot{s}=0$ is maintained through the entire stage for which $\Gamma >0$. 
In other words, the system may always be regarded as a perfect fluid [see the comment following Eq. (\ref{29})]. 
This implies that the variable $\tau $, which plays the role of a relaxation time here as well, has to be interpreted now in a different manner.  
While the quantity $\tau $ of the previous section (Sec. II) describes the relaxation from states with $\dot{s}>0$ to those with $\dot{s}=0$ 
due to internal (elastic) collision processes and is of the order of the mean free collision time, we have now a ``relaxation'' from states with 
$\dot{s}=0$ and $\Gamma >0$ to states with $\dot{s}=0$ and $\Gamma =0$. 
The corresponding process is {\it not} related to interactions in which the mean free collision time plays a role. 
Consequently, it will no longer be contradicting to have a relaxation time of the order of the Hubble time and, on the other hand, internal collision time scales which are small enough for a (perfect) fluid description to be sensible.   
This picture relies on a separation of interactions which guarantee a perfect fluid behavior from those which are responsible for the particle creation. 
The time scales of interest here refer to the latter, i.e., to the particle number non-preserving part of the interactions. 
The time scales of the former are assumed to be much smaller and will not explicitly appear in the following. 
With this basic difference in mind we will continue to use the symbol $\tau $ to characterize the ``relaxation'' from $\Gamma >0$ to $\Gamma =0$.    
          
Under the condition of ``isentropic'' particle production, the change rates for the relevant thermodynamic variables are 
\begin{equation}
\frac{\dot{n}}{n} = - \left(\Theta - \Gamma  \right)\ , 
\mbox{\ }\mbox{\ }\mbox{\ }
\mbox{\ }\mbox{\ }\mbox{\ }
\frac{\dot{T}}{T} = - \left(\Theta - \Gamma  \right)
\frac{\partial{p}}{\partial{\rho }}
\label{34}
\end{equation}
and 
\begin{equation}
\dot{\rho } = - \left(\Theta - \Gamma  \right)\left(\rho + p \right)\ , 
\mbox{\ }\mbox{\ }\mbox{\ }\mbox{\ }\mbox{\ }\mbox{\ }
\dot{p} = - c _{s}^{2} \left(\Theta - \Gamma  \right)\left(\rho + p \right)\ .
\label{35}
\end{equation}
With the help of these relations the entropy production density 
(\ref{33}) may be written as 
\begin{equation}
T S ^{m}_{;m} = nsT \Gamma  - 
\left[\left(\rho + p \right)\frac{\Gamma }{\Theta } \right]^{2}
\frac{\tau }{\zeta }
\left[\frac{\left(\Gamma / \Theta  \right)^{\displaystyle \cdot}}
{\Gamma / \Theta } 
- \left(\Theta - \Gamma  \right)\left(1 + c _{s}^{2} \right)
+ \frac{1}{2}
\left(\frac{\left(\tau / \zeta  \right)^{\displaystyle \cdot}}{\tau / \zeta } 
- \frac{\dot{T}}{T} + \Theta 
 \right)\right]\ .
\label{36}
\end{equation}
Inserting here for the last term 
\begin{equation}
\frac{\left(\tau / \zeta  \right)^{\displaystyle \cdot}}{\tau / \zeta } 
- \frac{\dot{T}}{T} + \Theta 
= \Theta \left(2 + c _{s}^{2} + \frac{\partial{p}}{\partial{\rho }} \right) 
+ \frac{\pi }{\rho + p}\left(1 + c _{s}^{2} 
+ \frac{\partial{p}}{\partial{\rho }} \right)\Theta 
- \frac{\left(c _{b}^{2} \right)^{\displaystyle \cdot}}{c _{b}^{2}}\ ,
\label{37}
\end{equation}
which follows with the help of Eqs. (\ref{14}), (\ref{34}), and (\ref{35}),    
the expression (\ref{36}) becomes 
\begin{equation}
TS ^{m}_{;m} = \left(\rho + p \right)\frac{\Gamma }{\Theta }
\left\{\frac{nsT}{\rho + p}\Theta 
- \left(\rho + p \right)\frac{\Gamma }{\Theta }\frac{\tau }{2 \zeta }
\left[\Gamma + 2 \frac{\left(\Gamma / \Theta  \right)^{\displaystyle \cdot}}
{\Gamma / \Theta } - \left(\Theta - \Gamma  \right)
\left(c _{s}^{2} - \frac{\partial{p}}{\partial{\rho }} \right) 
- \frac{\left(c _{b}^{2} \right)^{\displaystyle \cdot}}
{c _{b}^{2}  } \right]\right\}\ .
\label{38}
\end{equation}
Analogously to the standard procedure of Sec. II, the simplest way to guarantee $S ^{m}_{;m} \geq 0$ is again 
a generalized linear relation 
\begin{equation}
\left(\rho + p \right)\frac{\Gamma }{\Theta } = \zeta 
\left\{\frac{nsT}{\rho + p}\Theta 
- \left(\rho + p \right)\frac{\Gamma }{\Theta }\frac{\tau }{2 \zeta }
\left[\Gamma + 2 \frac{\left(\Gamma / \Theta  \right)^{\displaystyle \cdot}}
{\Gamma / \Theta } - \left(\Theta - \Gamma  \right)
\left(c _{s}^{2} - \frac{\partial{p}}{\partial{\rho }} \right) 
- \frac{\left(c _{b}^{2} \right)^{\displaystyle \cdot}}
{c _{b}^{2}  } \right]\right\}\ ,
\label{39}
\end{equation}
with $\zeta > 0$, which implies 
\begin{equation}
S ^{m}_{;m} = \frac{\left[\left(\rho + p \right) 
\frac{\Gamma }{\Theta }\right]^{2}}{\zeta T} \geq 0
\label{40}
\end{equation}
for the entropy production density.  
Use of Eq. (\ref{14})  in the generalized linear relation (\ref{39})   
allows us to obtain a dynamical equation for 
$\Gamma / \Theta $: 
\begin{equation}
\tau \left(\frac{\Gamma }{\Theta } \right)^{\displaystyle \cdot} 
+ \frac{\Gamma }{\Theta } = \Theta \tau 
\left[\frac{nsT}{\rho + p}c _{b}^{2} 
- \frac{1}{2}\left(1 + c _{s}^{2} - \frac{\partial{p}}{\partial{\rho }}\right) \left(\frac{\Gamma }{\Theta } \right)^{2} 
+ \frac{1}{2}\left(c _{s}^{2} - \frac{\partial{p}}{\partial{\rho }} \right)\frac{\Gamma }{\Theta } 
+ \frac{1}{2}\frac{\left(c _{b}^{2} \right)^{\displaystyle \cdot}}
{\Theta c _{b}^{2}}\frac{\Gamma }{\Theta }\right]\ . 
\label{41}
\end{equation} 
The same equation in terms of $\pi $ is  
\begin{equation}
\tau \dot{\pi } + \pi = - \rho \Theta \tau 
\left[\gamma c _{b}^{2}\frac{nTs}{\rho + p} 
+ \frac{\pi }{2 \rho }
\left(2 + c _{s}^{2} + \frac{\partial{p}}{\partial{\rho }} \right) 
+ \frac{\pi ^{2}}{2 \rho ^{2}}
\left(1 + c _{s}^{2} + \frac{\partial{p}}{\partial{\rho }} \right) \right]
+ \frac{\pi \tau }{2}
\frac{\left(c _{b}^{2} \right)^{\displaystyle \cdot}}
{ c _{b}^{2}}\ ,
\label{42}
\end{equation}
which is the counterpart of Eq. (\ref{12})  of Sec. II. 
For $\tau \rightarrow 0$ one recovers the Eckart-type relations 
(\ref{32}).  

The field equations (\ref{16})  and Eq. (\ref{29})  may be combined to yield 
\begin{equation}
\frac{\Gamma }{\Theta } = 1 + \frac{2}{3 \gamma }
\frac{\dot{H}}{H ^{2}}\ .
\label{43}
\end{equation}
Use of the last relation in Eq. (\ref{41})  provides us with  the following second-order equation for $H$,  
\begin{eqnarray}
&&\tau H \left[\frac{\ddot{H}}{H} 
- \frac{\dot{H}^{2}}{\gamma H ^{2}}
\left(1 + c _{s}^{2}  
+ \frac{\partial{p}}{\partial{\rho }}\right)  
+ 3\dot{H}\left(1 
- \frac{1}{2}\left(\frac{\partial{p}}{\partial{\rho }} 
- c _{s}^{2} \right)
\right)  
\right.
\nonumber\\
&&- \left. \frac{9}{2}H ^{2}\gamma 
\left(c _{b}^{2}\frac{nTs}{\rho + p} - \frac{1}{2}\right)
- \frac{1}{2}\frac{\left(c _{b}^{2} \right)^{\displaystyle \cdot}}
{H c _{b}^{2}}\left(\dot{H} + \frac{3}{2}\gamma H ^{2}\right)
\right] + \dot{H}
+ \frac{3}{2}\gamma H^{2} = 0 \ ,  
\label{44}
\end{eqnarray}
which is the counterpart of Eq. (\ref{19}) of Sec. II.  
For $p = \rho /3$ and $ns=\left(\rho +p \right)/T$ 
the general equation (\ref{44}) for $H$ reduces to 
\begin{equation}
\tau H \left[\frac{\ddot{H}}{H} - \frac{5}{4}\frac{\dot{H}^{2}}{H ^{2}} 
+ 3 \dot{H} - 6 H ^{2}\left(c _{b}^{2} - \frac{1}{2} \right)\right] 
+ \dot{H} + 2 H ^{2} = 0 
\mbox{\ }\mbox{\ }\mbox{\ }
\mbox{\ }\mbox{\ }\mbox{\ }
\mbox{\ \ \ \ }
\left(p = \frac{\rho }{3} \right)\ ,
\label{45}
\end{equation}
which is the analogue of  Eq. (\ref{20}). 
Again there exist stationary solutions $\dot{H} = 0$. 
From the relationship (\ref{43}) we have $\Gamma = 3H$ in this case, equivalent to constant values of $n$, $T$, $\rho $ and $p$ according to the evolution laws (\ref{34}) and (\ref{35}). 
The apparently unphysical behaviour of the temperature of the previous section (see the discussion following Eq. (\ref{21})) does no longer occur. 
Since $n$ is stationary as well, there is no exponential dilution of the fluid either. 
This behavior demonstrates that the extension of the fluid dynamics to not necessarily small values of $\pi $ is more satisfactory in case $\pi $ represents cosmological particle production.   
For the parameter $\tau H$  we find 
\begin{equation}
\tau H = \frac{1}{3}\frac{1}{c _{b}^{2} - \frac{1}{2}}\ .
\label{46}
\end{equation}
Comparing the last expression with Eq. (\ref{21}) for $B=C=0$ shows that particle production leads to a higher value for $\tau H$ than ``conventional'' causal cosmology, i.e., the relaxation time is enlarged.  
(We recall that the ``relaxation'' time $\tau $ here refers to a different process than in Sec.II). 
On the other hand, the applicability of our formalism is restricted to 
$c _{b}^{2} > \frac{1}{2}$. 
The general causality restriction $c _{b}^{2} \leq 1 - c _{s}^{2} \leq \frac{2}{3}$ [cf. Eq. (\ref{15})] is compatible with $c _{b}^{2} > \frac{1}{2}$. 
For values $c _{b}^{2} < \frac{1}{2}$ there does not exist a physically meaningful stationary solution of Eq. (\ref{44}).  
In the following we shall study the relativistic matter  dynamics for different cases on the basis of Eq. (\ref{45}) .  

\subsection{The linear case $\Gamma \propto H$ }
Looking for power-law solutions 
$a \propto t ^{q}$ of Eq. (\ref{45})  one obtains
\begin{equation}
\tau H = \frac{2}{3} \frac{q \left(q - \frac{1}{2} \right)}
{q \left[\left(1 + 2q \left(c _{b}^{2}  - \frac{1}{2} \right) \right) \right] 
- \frac{1}{4}}\ .
\label{47}
\end{equation}
The particle production rate becomes
\begin{equation}
\Gamma = \left(1 - \frac{1}{2q}\right)3H \ ,
\label{48}
\end{equation}
i.e., $\Gamma \propto H$ follows as a consequence of the power-law ansatz, while the expression (\ref{24})  remains valid with $\pi $ given by 
Eq. (\ref{29}).   
This allows us to write the parameter $\tau H$   as 
\begin{equation}
\tau H = \frac{1}{3}
\left(1 - \frac{1}{2q} \right)
\frac{1}{c _{b}^{2} 
- \frac{1}{2}\left(1 - \frac{1}{2q} \right)^{2}}\ ,
\label{49}
\end{equation}
or, in terms of $\Gamma $, 
\begin{equation}
\tau H = \frac{\Gamma }{9H }
\frac{1}{c _{b}^{2} - \frac{1}{2}
\left(\frac{\Gamma }{3H } \right)^{2}}\ .
\label{50}
\end{equation}
For $\Gamma \rightarrow 3H $, corresponding to 
$q \rightarrow \infty$ we recover the case (\ref{46}) . 
Comparison of the parameters (\ref{49})  and (\ref{22})  for 
$\gamma =4/3$ and $Q=1$ shows that also 
in the present case the relaxation is generally larger than for conserved particle numbers. 
Since $\Theta - \Gamma $ is 
\begin{equation}
\Theta - \Gamma = \frac{3}{2q}H \ ,
\label{51}
\end{equation}
the balances (\ref{34}) and (\ref{35}) reduce to    
\begin{equation}
\frac{\dot{n}}{n} = - \frac{3}{2q}H \ , 
\mbox{\ }\mbox{\ }\mbox{\ }
\mbox{\ }\mbox{\ }\mbox{\ }
\frac{\dot{T}}{T} = - \frac{1}{2q}H \ , 
\mbox{\ }\mbox{\ }\mbox{\ }
\mbox{\ }\mbox{\ }\mbox{\ }
\frac{\dot{\rho }}{\rho } = - \frac{2}{q}H \ ,
\label{52}
\end{equation}
resulting in 
\begin{equation}
n \propto a ^{-\frac{3}{2q}}\ , 
\mbox{\ }\mbox{\ }\mbox{\ }
\mbox{\ }\mbox{\ }\mbox{\ }
T \propto a ^{-\frac{1}{2q}}\ , 
\mbox{\ }\mbox{\ }\mbox{\ }
\mbox{\ }\mbox{\ }\mbox{\ }
\rho \propto a ^{- \frac{2}{q}}\ .
\label{53}
\end{equation}
For $q = \frac{1}{2}$ we recover the standard perfect fluid behaviour. 
For $q > \frac{1}{2}$, equivalent to $\Gamma > 0$ all quantities decrease at lower rates than for $q = \frac{1}{2}$. 
Unlike the $\Gamma = 0$ case of Sec. II, the equilibrium 
relation $n \propto T ^{3}$ continues to hold here for any value of $q$. 
An essential advantage of the present setting is an inherent self-limitation 
of the theory, connected with the denominator 
$c _{b}^{2}- \frac{1}{2}\left(\frac{\Gamma }{\Theta } \right)^{2} = 
c _{b}^{2} -\frac{9}{32}\left(\frac{\pi }{\rho } \right)^{2}$ in Eq. (\ref{50}). 
Our considerations make only sense for 
$\left(\frac{\pi }{\rho } \right)^{2} < \frac{32}{9} c _{b}^{2}$. 
Otherwise we would obtain a negative $\tau $ which is clearly unphysical. 
This represents an example for the property mentioned below Eq. (\ref{33}), 
according to which the magnitude of the effective viscous pressure may be limited by the internal fluid dynamics.

\subsection{The quadratic case $\Gamma \propto H ^{2}$ }
To study more general situations it is convenient to replace $\dot{H}$ in 
Eq. (\ref{43})  by  
\[
\dot{H} = \frac{\mbox{d}H}{\mbox{d}a}\dot{a} = H ^{\prime }H a \ ,
\]
where $H ^{\prime } \equiv  \mbox{d}H/ \mbox{d}a$. 
The resulting equation is 
\begin{equation}
\frac{H ^{\prime }}{H \left[\frac{\Gamma }{3H} - 1 \right]} 
= \frac{3}{2}\frac{\gamma }{a}\ .
\label{56}
\end{equation}
For $\Gamma = \alpha 3H$ with $\alpha = {\rm const}$, it integrates to 
\begin{equation}
\frac{H}{H _{0}} 
= \left(\frac{a _{0}}{a}\right)^{\frac{3}{2}\gamma \left(1- \alpha \right)}
\mbox{\ }\mbox{\ }\mbox{\ }
\mbox{\ }\mbox{\ }\mbox{\ }
\mbox{\ }\mbox{\ }\mbox{\ }
\mbox{\ }\mbox{\ }\mbox{\ }
\left(\Gamma = \alpha 3 H \right)\ ,
\label{57}
\end{equation}
and 
\begin{equation}
a \propto t ^{\frac{2}{3 \gamma \left(1- \alpha \right)}}
\mbox{\ }\mbox{\ }\mbox{\ }
\mbox{\ }\mbox{\ }\mbox{\ }
\mbox{\ }\mbox{\ }\mbox{\ }
\mbox{\ }\mbox{\ }\mbox{\ }
\left(\Gamma = \alpha 3 H \right)\ ,
\label{58}
\end{equation}
which are the previous results for the linear case:  
A linear relation between $\Gamma $ and $H$ implies a power-law behaviour of the scale factor and vice versa.   
A more interesting case is 
a dependence $\Gamma \propto H ^{2}$, which was already discussed in \cite{GunzMaNe} in the context of matter creation due to a decaying cosmological constant:
\begin{equation}
\frac{\Gamma }{3H} = \beta \frac{H}{H _{e}}\ ,
\label{59}
\end{equation}
with $\beta = {\rm const}$. The quantity $H _{e} = H \left(a _{e} \right)$ is the Hubble parameter at some fixed epoch with $a = a _{e}$. 
With the quadratic ansatz (\ref{59})  integration of the differential equation (\ref{56})  yields 
\begin{equation}
H = \frac{a _{e} ^{\frac{3}{2}\gamma }}
{\left(1 - \beta \right)a ^{\frac{3}{2}\gamma } 
+ \beta a _{e} ^{\frac{3}{2}\gamma }}H _{e}\ .
\label{60}
\end{equation}
For $a \rightarrow 0$ we have $H \rightarrow \beta ^{-1}H _{e}={\rm const}$, i.e., an exponentially accelerated expansion ($\ddot{a}>0$), while for 
$a \gg a _{e}$ we recover the familiar dependence 
$H \propto a ^{-3 \gamma /2}$ for FLRW universes (with 
$\ddot{a}<0$). 
If we identify the intermediate value of $a$ at which $\ddot{a}=0$ with 
$a _{e}$, we have $\dot{H}_{e} = - H _{e}^{2}$. 
From Eqs. (\ref{43})  and (\ref{59})  it follows that $\beta $ is then fixed according to 
\begin{equation}
\beta = 1 - \frac{2}{3 \gamma }\ ,
\label{61}
\end{equation}
i.e., for relativistic matter we find  
\begin{equation}
\beta = \frac{1}{2} 
\quad\Rightarrow\quad 
H = 2\frac{a _{e} ^{2}}
{a ^{2} + a _{e} ^{2}}H _{e}
\ .
\label{62}
\end{equation}
While a linear dependence of $\Gamma $ on $H$ corresponds to a power-law solution for the scale factor, we have now 
\begin{equation}
t = t _{e} + \frac{1}{4H _{e}}
\left[\ln \left(\frac{a}{a _{e}} \right)^{2} + 
\left(\frac{a}{a _{e}} \right)^{2} - 1\right]\ .
\label{63}
\end{equation}
This is the same dependence as the one obtained on the basis of a phenomenological model for a decaying cosmological constant by Gunzig et al. \cite{GunzMaNe}. (For alternative thermodynamic descriptions of cosmological vacuum decay see \cite{Limavac,JeGer}). 
The resulting cosmological scenario, however, is  different here from the model in \cite{GunzMaNe}. In our case $\Gamma $ is limited to $\Gamma \leq 3H$, while in \cite{GunzMaNe} the creation rate may be larger than this value. 
This gives rise to a difference in the thermal evolution at very early stages (see below). 
But our main point here is the following. 
In obtaining the solution (\ref{62}) we have not really solved our basic equation (\ref{44}) but only Eq. (\ref{43}) under the assumption (\ref{59}). 
The remaining part of the dynamics is given by Eq. (\ref{41}). 
In other words, we are dealing with an enlarged set of basic equations.    
For $p = \rho /3$ and vanishing chemical potential 
Eq. (\ref{41}) may be written as  
\begin{equation}
\Theta \tau = \frac{\frac{\Gamma }{\Theta }}
{c _{b}^{2} 
- \frac{1}{2} \left(\frac{\Gamma }{\Theta } \right)^{2} 
- \frac{\left(\Gamma / \Theta  \right)^{\displaystyle \cdot}}{\Theta }
}\ .
\label{55}
\end{equation}
It is obvious that the dynamical degree of freedom 
$\Gamma / \Theta $ determines the parameter $\Theta \tau $ as well. 
Since we prescribed a specific functional dependence for $\Gamma /H $, Eq. (\ref{55}) fixes the parameter 
$\tau H$. 
On the other hand, in order to be physically meaningful (positive relaxation time), any choice of $\Gamma / \Theta $ is restricted to 
\begin{equation}
\frac{1}{2} \left(\frac{\Gamma }{\Theta } \right)^{2} 
+ \frac{\left(\Gamma / \Theta  \right)^{\displaystyle \cdot}}{\Theta } 
< c _{b}^{2}  \ .
\label{55a}
\end{equation}
This condition on $\Gamma $ (or, equivalently, on the corresponding effective viscous pressure according to Eq. (\ref{29})) is a new result.  
In the present case 
the left-hand side of the last relation becomes 
\begin{equation}
\frac{1}{2} \left(\frac{\Gamma }{\Theta } \right)^{2} 
+ \frac{\left(\Gamma / \Theta  \right)^{\displaystyle \cdot}}{\Theta }
= - \frac{1}{3}\frac{H}{H _{e}}
\left[1 - \frac{7}{8}\frac{H}{H _{e}} \right]
\ .
\label{64}
\end{equation}
It follows that the parameter $\tau H$ is given by  
\begin{equation}
\tau H = \frac{1}{6}\frac{1}
{c _{b}^{2} 
+ \frac{1}{3}\frac{H}{H _{e}}
\left[1 - \frac{7}{8}\frac{H}{H _{e}} \right]}
\frac{H}{H _{e}}\ .
\label{65}
\end{equation}
The condition (\ref{55a}) specifies to 
\begin{equation}
H < \frac{4}{7}\left[1 + \sqrt{1 + \frac{21}{2}c _{b}^{2}} \right]H _{e}\ .
\label{66}
\end{equation}
This represents a restriction of the admissible Hubble rate in dependence of the parameter $c _{b}^{2}$.  
Since $H$ starts at $H _{0} = 2H _{e}$
[cf. Eq. (\ref{62}) for $a \rightarrow 0$], this inequality is always satisfied 
for $\frac{2}{3}\geq c _{b}^{2} > \frac{1}{2}$, i.e., our whole scenario is compatible with general causality requirements for effective bulk viscous fluids from the beginning.  
For the maximally acceptable value of $c _{b}^{2}$ which is compatible with the causality requirement, $c _{b}^{2} = 2/3$,  
(see the discussion following Eq. (\ref{46})), the initial value of $\tau $ is  
\begin{equation}
\tau _{0} = 2 H _{0}^{-1}
\mbox{\ }\mbox{\ }\mbox{\ }
\mbox{\ }\mbox{\ }\mbox{\ }
\mbox{\ }\mbox{\ }\mbox{\ }
\mbox{\ }\mbox{\ }\mbox{\ }
\mbox{\ }\mbox{\ }\mbox{\ }
\left(c _{b}^{2}  = \frac{2}{3} \right)\ ,
\label{67}
\end{equation}
i.e., the relaxation time exceeds the Hubble expansion time. 
For $c _{b}^{2}<\frac{1}{2}$ the theory is only applicable at later periods of the evolution but not at very early times. 
Even for $c _{b}^{2}  \ll 1$, however, where condition (\ref{66}) becomes 
$H<\frac{8}{7}H _{e}$, the applicability still starts within the phase of accelerated expansion ($H>H _{e}$). 
These restrictions are again manifestations of the fact that the theory considered here provides its own limits of validity. 
At the same time this precises the conditions under which a fluid picture of the inflationary universe is permitted.     

The results obtained so far give rise to the following conclusions. 
As long as one assumes a power-law solution for the scale factor, $\Gamma $ is necessarily proportional to $H$, i.e., the ratio 
$\Gamma /H$ is constant. 
The impact of a nonvanishing particle production rate on the cosmological dynamics does not change in time under this condition. 
Given $\Gamma $ is large enough at a certain time to generate a power  
$q>1$, i.e., accelerated expansion, there is no way to terminate such a phase. 
Consequently, models with $a \propto t ^{q}$ which imply 
$\Gamma /H = {\rm const}$ may be realistic at most piecewise. 
For a model with $\Gamma \propto H ^{2}$, on the other hand, the time dependence of the scale factor is more involved (see Eq. (\ref{63})) but there is a variation in the ratio $\Gamma /H$. 
The impact of a nonvanishing particle production rate on the cosmological dynamics is different in different epochs. 
In particular, it implies a transition from an inflationary phase to a conventional, i.e. decelerating FLRW period. 
This transition is accompanied by a decrease of the parameter $\tau H$. 
For radiation the latter quantity is generally given by Eq. (\ref{65}) . 
For $H \ll H _{e}$, equivalent to $a \gg a _{e}$ the ratio $\tau H$  tends to zero and the cosmic substratum approaches a perfect relativistic fluid behavior with $\Gamma /H \ll 1$, i.e., negligible particle production. 

Combination of the Friedmann equation (\ref{16}) with the Hubble parameter (\ref{62})  allows us to obtain 
$\rho = \rho \left(a \right)$, 
\begin{equation}
\rho = \frac{3 H _{e}^{2}}{2 \pi }m _{p}^{2}
\left[\frac{a _{e}^{2}}{a ^{2} + a _{e}^{2}} \right]^{2}\ ,
\label{68}
\end{equation}
where we have replaced $\kappa$ by the Planck mass $m _{p}$  according to  $\kappa = 8 \pi /m _{p}^{2}$. 
In the following we shall assume $\frac{2}{3}\geq c _{b}^{2} > \frac{1}{2}$ to ensure that condition (\ref{66}) is always satisfied.  
Initially, i.e., for $a \rightarrow 0$, corresponding to 
$t \rightarrow -\infty$, the energy density is constant, while for 
$a \gg a _{e}$ the familiar behaviour 
$\rho \propto a ^{-4}$ is recovered. 
In our approach the evolution of the universe starts in a quasistationary state with a finite initial energy density, while  
the two-component model of Gunzig et al. \cite{GunzMaNe} implies a vanishing radiative energy density at $a=0$ for $t \rightarrow - \infty$. 
This  apparent contradiction is easily resolved, however, if we split the energy density (\ref{68}) according to $\rho = \rho _{1} + \rho _{2}$ with 
\[
\rho _{1} = \frac{3 H _{e}^{2}}{2 \pi }m _{p}^{2}
\left(\frac{a}{a _{e}} \right)^{2}
\left[\frac{a _{e}^{2}}{a ^{2} + a _{e}^{2}} \right]^{3}\ , 
\mbox{\ }\mbox{\ }\mbox{\ }
\rho _{2} = \frac{3 H _{e}^{2}}{2 \pi }m _{p}^{2}
\left[\frac{a _{e}^{2}}{a ^{2} + a _{e}^{2}} \right]^{3}\ .
\]     
The part $\rho _{2}$ corresponds to a cosmological ``constant'' in \cite{GunzMaNe} which decays as $a ^{-6}$ for $a \gg a _{e}$, while 
the part $\rho _{1}$ describes relativistic matter with 
$\rho _{1}\rightarrow 0$ for $a \rightarrow 0$. 
While this makes obvious that an interpretation of our approach in terms of a decaying cosmological ``constant'' is also admitted, the dynamics of particle number density and temperature is different from \cite{GunzMaNe}.  
In our case these quantities behave as 
\begin{equation}
n = n _{0}
\left[\frac{a _{e}^{2}}{a ^{2} + a _{e}^{2}} \right]^{3/2} 
= 2 \sqrt{2} n _{e}
\left[\frac{a _{e}^{2}}{a ^{2} + a _{e}^{2}} \right]^{3/2} 
\label{69}
\end{equation}
and
\begin{equation}
T = T _{0}
\left[\frac{a _{e}^{2}}{a ^{2} + a _{e}^{2}} \right]^{1/2} 
= \sqrt{2} T _{e}
\left[\frac{a _{e}^{2}}{a ^{2} + a _{e}^{2}} \right]^{1/2} \ ,
\label{70}
\end{equation}
respectively, where the subscripts $0$  and $e$ refer to the corresponding values at $a=0$ and 
at $a=a _{e}$. 
All the thermodynamic variables remain finite for $a \rightarrow 0$ 
and they approach the correct dependences for radiation for 
$a \gg a _{e}$. 
In particular, the universe starts with a finite maximal temperature, while the initial radiation temperature in \cite{GunzMaNe} is zero. 
In a sense, the radiative matter itself plays the role of a decaying cosmological ``constant''  in our approach. 
During the interval 
$- \infty < t \leq t _{e}$, equivalent to 
$0 < a \leq a _{e}$, the energy density decreases by a factor of $4$, while $n$ and $T$ decrease by factors of $2 \sqrt{2}$ and $\sqrt{2}$, respectively. 
For the number $N = n a ^{3}$ of particles in a comoving volume $a ^{3}$ we find 
\begin{equation}
N = 2 \sqrt{2} N _{e}
\left[\frac{a ^{2}}{a ^{2} + a _{e}^{2}} \right]^{3/2} 
= N _{f}
\left[\frac{a ^{2}}{a ^{2} + a _{e}^{2}} \right]^{3/2} \ ,
\label{71}
\end{equation}
where $N _{e} = n _{e}a _{e}^{3}$. 
The particle number $N$ starts from $N _{0}=0$  
at $a = 0$ and subsequently grows to the final constant value $N _{f} = 2 \sqrt{2}N _{e}$ for $a \gg a _{e}$. This value may be identified with the number of particles in the presently observed universe which is of the order $10 ^{88}$. 
Since the epoch $a = a _{e}$ with $\ddot{a}\left(t _{e} \right)=0$ the total particle number has grown by a factor of $2 \sqrt{2} $. 
The circumstance that particles are absent initially ($N _{0}=0$) is consistent with the interpretation of the corresponding initial energy density 
\begin{equation}
\rho _{0} = \frac{3 H _{e}}{2 \pi }H _{e}^{2}\ ,
\label{72}
\end{equation}
following from Eq. (\ref{68}) in the limit $a \rightarrow 0$,  
as a vacuum energy density. 
Consequently, 
our approach may be regarded as equivalent to the model of a decaying vacuum.   

We finish this section by considering the behavior of the entropy in our model universe. 
With Eqs. (\ref{14}), (\ref{29}), (\ref{59}), and (\ref{62}), the entropy flow vector (\ref{5})  may be written as 
\begin{equation}
S ^{a} = n s ^{*}u ^{a}\ ,
\label{73}
\end{equation}
where $s ^{*}$ is   
\begin{equation}
s ^{*} \equiv   \left[1 - \frac{1}{2 c _{b}^{2}}
\left(\frac{1}{2} \frac{H}{H _{e}}\right)^{2} \right]s 
= \left[1 - \frac{1}{2 c _{b}^{2}}
\left(\frac{a _{e}^{2}}{a ^{2} + a _{e}^{2}}\right)^{2} \right]s \ .
\label{74}
\end{equation}
This quantity is an analogue to the non-equilibrium entropy per particle for viscous fluids with conserved particle number. 
In the present context it is more adequately called an effective entropy per particle. 
For $a \rightarrow 0$ we have (recall that $s=4$) 
\begin{equation}
s ^{*}_{0}  
= 4 \left[1 - \frac{1}{2 c _{b}^{2}} \right]\ .
\label{75}
\end{equation}
This quantity is positive for $c _{b}^{2} > 1/2$ which coincides with the condition for the relaxation time to be positive 
[cf. Eq. (\ref{65}) for $H=H _{0}=2H _{e}$ ]. 
With $c _{b}^{2} =2/3$, the maximally allowable value, we get 
\begin{equation}
s ^{*}_{0} = 1 
\mbox{\ }\mbox{\ }\mbox{\ }
\mbox{\ }\mbox{\ }\mbox{\ }
\mbox{\ }\mbox{\ }\mbox{\ }
\mbox{\ }\mbox{\ }\mbox{\ }
\mbox{\ }\mbox{\ }\mbox{\ }
\mbox{\ }\mbox{\ }\mbox{\ }
\left(c _{b}^{2} =\frac{2}{3} \right)\ .
\label{76}
\end{equation}
The effective  entropy $s ^{*}$ per particle is reduced considerably compared with the value $s=4$ for $\Gamma =0$ . 
A reduction to $s ^{*}=1$ is compatible with the causality requirement. 
During the period where the particle production is highest, i.e., for 
$H > H _{e}$, the effective ``nonequilibrium'' contribution to the entropy is highest as well. 
For $H \ll H _{e}$ corresponding to $a \gg a _{e}$, the quantity  $s ^{*}$ 
approaches $s$. 
While $s$ is constant according to condition (\ref{29}), $s ^{*}$ is not. 
From the definition (\ref{74}) we obtain 
\begin{equation}
\dot{s} ^{*} = \frac{16}{c _{b}^{2}  }
\frac{a ^{2}}{a ^{2} + a _{e}^{2}}
\left[\frac{a _{e}^{2}}{a ^{2} + a _{e}^{2}} \right]^{3}H _{e}\ .
\label{77}
\end{equation}
This expression grows as 
$\dot{s}^{*} \propto a ^{2}$ for $a \ll a _{e}$ and decays as 
$\dot{s}^{*} \propto a ^{-6}$ for $a \gg a _{e}$. 
The effective entropy in a comoving volume is given by 
\begin{equation}
S ^{*} \equiv  n s ^{*}a ^{3} = N s ^{*}\ .
\label{78}
\end{equation}
It increases both due to the increase in $N$ from $N _{0}=0$ to $N _{f}$ and due to the increase of $s ^{*}$ to $s ^{*}_{f}=s=4$ for 
$a \gg a _{e}$. 
We emphasize that different from ``standard'' inflationary models which rely on the dynamics of a scalar field, this scenario does not imply a separate reheating phase. 
The mechanism which drives inflation is at the same time entropy producing. 
In the following Section we establish a relation between the fluid model of the universe presented so far and models which are based on the dynamics of scalar fields. 

\section{Scalar field dynamics}
Inflationary cosmology is usually discussed in terms of scalar fields with suitable potentials. 
The usual picture consists of an initial  
``slow roll'' phase with a dynamically dominating (approximately constant) potential term, which generates a de Sitter like exponential expansion, connected with an ``adiabatic supercooling''. 
During the subsequent ``reheating'' phase which in itself is of a highly complicated non-equilibrium structure, the scalar field is assumed to decay into ``conventional'' matter. 
The entire entropy in the presently observable universe is produced during the reheating phase according to these scenarios. 
As to the mechanism of entropy production the ``standard'' (i.e. based on scalar field dynamics) inflationary picture essentially differs from the fluid particle production scenario of the present paper, since here the entropy is produced already during the period of accelerated expansion. 
On the other hand, as far as the dynamics of the scale factor is concerned, there exists a close correspondence 
between the role of an effective negative fluid pressure (due to the production of particles) and a suitable scalar field potential. 
What counts here is the magnitude of the effective negative pressure, independently of its origin (scalar field potential or particle production).  
What one would like to have is an interconnection between both these different lines of describing the early universe. This would allow us to switch from the scalar field to the fluid picture and vice versa. 
It is the purpose of this Section to show  how the effective imperfect fluid dynamics discussed so far in this paper may be translated into the dynamics of a minimally coupled scalar field.  
In particular, we shall find a scalar field equivalent for the previously considered [cf. Sec. III] comoving fluid entropy which was shown to be increasing in time due to cosmological particle creation. 
To this purpose we start with the familiar identifications 
\begin{equation}
\rho = \frac{1}{2}\dot{\phi }^{2} + V \left(\phi  \right) \ , 
\mbox{\ \ \ \ \ \ \ } 
P \equiv  p + \pi 
= \frac{1}{2}\dot{\phi }^{2} - V \left(\phi  \right)\ ,
\label{79}
\end{equation}
or, 
\begin{equation}
\rho + P = \dot{\phi }^{2} \ , 
\mbox{\ \ \ \ \ \ \ \ } 
\rho - P = 2 V \left(\phi  \right)\ .
\label{80}
\end{equation}
With the expressions (\ref{68}), (\ref{29}), (\ref{59}), and (\ref{62}),  we obtain for $\rho + P$ 
\begin{equation}
\rho + P = 4\frac{H _{e}^{2}}{2 \pi }m _{p}^{2}
\frac{a ^{2}}{a _{e}^{2}}
\left[\frac{a _{e}^{2}}{a ^{2} + a _{e}^{2}} \right]^{3} 
= \dot{\phi }^{2}\ .
\label{81}
\end{equation}
For $\dot{\phi }$ we write 
$\dot{\phi }=\phi ^{\prime }\dot{a}=\phi ^{\prime }aH$, where 
$\phi ^{\prime }\equiv  \mbox{d}\phi / \mbox{d}a$. 
It follows that 
\begin{equation}
\phi ^{\prime } = \pm \frac{m _{p}}{2 \pi }
\frac{1}{\sqrt{a ^{2} + a _{e}^{2}} }\ .
\label{82}
\end{equation}
Integration of this equation yields 
\begin{equation}
\phi - \phi _{0} =  \frac{m _{p}}{\sqrt{2 \pi }}
{\rm Arsh} \frac{a}{a _{e}}
\quad\Rightarrow\quad 
\frac{a}{a _{e}} = \sinh \left[\sqrt{2 \pi }
\frac{\phi - \phi _{0}}{m _{p}}\right]\ ,
\label{83}
\end{equation}
where we have restricted ourselves to $\phi \geq \phi _{0}$.  
The potential is given by 
\begin{equation}
\frac{\rho - P}{2} = \frac{H _{e}^{2}}{2 \pi }m _{p}^{2}
\left[\frac{a ^{2}}{a _{e}^{2}} + 3 \right]
\left[\frac{a _{e}^{2}}{a ^{2} + a _{e}^{2}} \right]^{3} 
= V\ , 
\label{84}
\end{equation}
or, 
\begin{equation}
V = \frac{H _{e}^{2}}{2 \pi }m _{p}^{2}
\frac{\cosh ^{2}\left[ \sqrt{2 \pi }
\frac{\phi - \phi _{0}}{m _{p}}\right] + 2}
{\cosh ^{6}\left[\sqrt{2 \pi }
\frac{\phi - \phi _{0}}{m _{p}}\right]} 
= \frac{H _{e}^{2}}{2 \pi }m _{p}^{2}
\frac{3 - 2 \tanh ^{2}\left[ \sqrt{2 \pi }
\frac{\phi - \phi _{0}}{m _{p}}\right]}
{\cosh ^{4}\left[\sqrt{2 \pi }
\frac{\phi - \phi _{0}}{m _{p}}\right]} 
\mbox{\ }\mbox{\ }\mbox{\ }
\mbox{\ }\mbox{\ }\mbox{\ }
\mbox{\ }\mbox{\ }\mbox{\ }
\mbox{\ }\mbox{\ }\mbox{\ }
\left(\Gamma = \frac{3}{2}\frac{H ^{2}}{H _{e}}  \right)
\ .
\label{85}
\end{equation}
This expression was found by Maartens et al. 
\cite{MaaTayRou} as exactly that potential which produces the dynamical behaviour (\ref{63}) of the scale factor.  
The rate $\Gamma $ as a function of $\phi $ is 
\begin{equation}
\Gamma = \frac{6H _{e}}{\cosh ^{4}\left[\sqrt{2 \pi }
\frac{\phi - \phi _{0}}{m _{p}}\right]}\ .
\label{86}
\end{equation}
In terms of the scalar field the cosmic evolution starts with 
\begin{equation}
V _{0} = \frac{3 H _{e}^{2}}{2 \pi }m _{p}^{2} 
\label{87}
\end{equation}
and $\dot{\phi }_{0}=0$. 
At $a=a _{e}$ the value of the potential is reduced to  
\begin{equation}
V _{e} = \dot{\phi }_{e}^{2} = \frac{H _{e}^{2}}{4 \pi }m _{p}^{2} \ .
\label{88}
\end{equation}
The relations found so far in this section allow us to change from the fluid to the scalar field picture and vice versa at any time of the cosmological evolution. 
In particular, they may be used to express the entropy (\ref{78}) in terms of the scalar field. Namely, combining Eqs. (\ref{71}), (\ref{74}) and (\ref{78}) with the second relation of Eq. (\ref{83}), we obtain 
\begin{equation}
S ^{*} = 4 N _{f}\tanh ^{3}\left[\sqrt{2 \pi }
\frac{\phi - \phi _{0}}{m _{p}}\right]
\left(1 - \frac{1}{c _{b}^{2}}\cosh ^{-4}\left[\sqrt{2 \pi }
\frac{\phi - \phi _{0}}{m _{p}}\right] \right)\ 
\label{88a}
\end{equation}
as the ``effective'' scalar field entropy of a comoving volume. 
This amount of entropy may be attributed to the universe, independently of the specific matter model. 
The corresponding first-order limit $S$ for $c _{b}^{2} \rightarrow \infty$ is  
\begin{equation}
S ^{*} \rightarrow S = 4 N _{f}\tanh ^{3}\left[\sqrt{2 \pi }
\frac{\phi - \phi _{0}}{m _{p}}\right]
\ .
\label{88b}
\end{equation}
The property of the scalar field configuration to represent a certain amount of entropy may be used to avoid solutions of the type of ``jump'' solutions \cite{MaaTayRou} according to which an instantaneous reheating due to a discontinuity in the effective pressure at the transition from an early scalar field epoch to a subsequent radiation dominated FLRW phase is supposed to 
generate the entropy in the universe. 

We mention that also for a power-law behaviour $a \propto t ^{q}$, equivalent to the linear dependence (\ref{48}),  an equivalent scalar field dynamics may be formulated. 
By similar steps like those used to derive the expression (\ref{84}),  we find for the potential in the linear case 
\begin{equation}
V = \frac{\left(3q - 1 \right)q}{8 \pi }
\frac{m _{p}^{2}}{t _{i}^{2}}
\exp{\left[4\sqrt{\frac{ \pi }{q}}
\frac{\phi - \phi _{i}}{m _{p}} \right]}
\mbox{\ }\mbox{\ }\mbox{\ }
\mbox{\ }\mbox{\ }\mbox{\ }
\mbox{\ }\mbox{\ }\mbox{\ }
\mbox{\ }\mbox{\ }\mbox{\ }
\left(\Gamma = \left(1 - \frac{1}{2q}\right)3H  \right)\ ,
\label{89}
\end{equation}
where $t _{i}$ is some initial reference time. 
The potentials (\ref{85}) and (\ref{89}) are of the type of exponential potentials. 
Exponential potentials  arise naturally in  particle physics \cite{SaZe,Halli} 
and are known to be typical for power-law solutions for the 
scale factor \cite{LuMa,JohnB,BuBa,Liddle}. They were also studied in the context of interacting scalar field cosmologies \cite{BiCol}
and in discussions of the recent SNIa results according to which the present universe might be accelerating \cite{HutTur,Matos}. 

The option to use either the fluid or the scalar field picture and to switch between them gives rise to a complementary, more comprehensive overall picture  of the dynamics of the universe. 
Moreover, it may contribute to establish a link between fluid cosmology and particle-physics-motivated investigations of the early universe. 

As a final remark we point out that rewriting the fluid dynamics into the dynamics of a scalar field implies only those subset of equations which 
does not depend on the parameter $\tau H$.   
There is no counterpart of relation (\ref{65}) in ``conventional'' scalar field theory. 
Compared with the latter, the causal fluid dynamics is more restrictive.  
It provides us with limits for the admissible inflationary dynamics which do not exist for scalar fields. 
Again, this may be traced back to the inherent self-limitation of the causal theory for isentropic particle production.

\section{Conclusion}
We have developed a setting according to which isentropic cosmological particle production is a dynamical process described with the help of the second-order Israel-Stewart theory for (effectively) imperfect fluids. 
This picture uses the well-known equivalence between an effective viscous fluid pressure and cosmological particle creation to make the rate for the latter a dynamical degree of freedom. 
The resulting dynamics was shown to be inherently self-limiting, i.e., it may restrict the magnitude of the effective viscous pressure. 
We have precised the conditions under which an effective bulk pressure,  
compatible with the general causality requirements of second-order thermodynamics of the Israel-Stewart type, may drive a  phase of inflationary expansion. 
In particular, for 
a quadratic dependence of the particle production rate on the Hubble parameter  
we have established a scenario according to which the evolution of the universe starts in a de Sitter phase with finite, stationary values of density and temperature, and subsequently evolves smoothly to a standard FLRW stage.  
The corresponding relaxation times turned out to be of the order of the Hubble time during the phase of accelerated expansion.  
This scenario was also shown to be consistent with the model of a vacuum energy which decays into relativistic matter. 
Furthermore, we have clarified that part of the effective imperfect fluid dynamics is equivalent to the dynamics of a minimally coupled scalar field.  
This correspondence was used to find a scalar field equivalent for the 
(generally increasing) comoving fluid entropy.

\acknowledgments
This paper was supported by the Deutsche Forschungsgemeinschaft.

\end{document}